\documentclass[preprint,prc,showpacs,preprintnumbers,amsmath,amssymb,floatfix]{revtex4}
\usepackage{amsmath}

\usepackage{graphicx}
\usepackage{dcolumn}
\usepackage{bm}
\usepackage{ulem} 
\usepackage[usenames]{color}
\usepackage{epstopdf}
\usepackage{epsfig}
\usepackage{float}
\usepackage{subfigure}
\usepackage{multirow}
\usepackage[version=3]{mhchem}

\newcommand{\nc}{\newcommand}       
\nc{\vc}[1] {\mbox{\boldmath $#1$}} 
\nc{\del}       {\partial}              
\nc{\bra}       {\langle}               
\nc{\ket}       {\rangle}               
\nc{\bras}[1]   {\langle #1|}           
\nc{\kets}[1]   {|#1\rangle}            
\nc{\mapleft}[1]{           
 \smash{\mathop{\,          %
  \hbox to 1.5cm{\rightarrowfill}\, }\limits_{#1}}}
\nc{\beq}     {\begin{eqnarray}} \nc{\eeq}    {\end{eqnarray}}
\nc{\nn}      {\\\nonumber} \nc{\vs}      {\vspace{-0.275cm}}
\nc{\fra}    {\frac{1}{2}}
\nc{\mb}        {\mathbf}

\begin{document}

\preprint{}

\title{Single-$\Xi^-$ hypernuclei with quark mean field model}

\author{Jinniu Hu}
\email{hujinniu@nankai.edu.cn}
\affiliation{School of Physics, Nankai University, Tianjin 300071,  China}
\author{Hong Shen}
\email{songtc@nankai.edu.cn}
\affiliation{School of Physics, Nankai University, Tianjin 300071,  China}

\date{\today}
\begin{abstract}
The single-$\Xi^-$ hypernuclei are studied using a quark mean field model. At the quark level, the $\Xi^-$ hyperon is composed of one $d$ quark and two $s$ quarks, which are confined by the harmonic oscillator potentials.  In the case of hadrons, the baryons interact with each other by exchanging $\sigma, ~\omega$, and $\rho$ mesons between quarks in different baryons. The single-$\Xi^-$ binding energies of $\Xi$ hypernuclei are investigated from $\ce{_{\Xi^{-}}^{12}Be}$ to $\ce{^{208}_{\Xi^-}Tl}$  using different parameter sets, which are determined by the ground-state properties of several stable nuclei and the empirical values of the single-$\Lambda$ and $\Xi$ potentials at the nuclear saturation density. For the bound states of $\ce{\Xi^{-}{+}^{14}N}$ (i.e., $\ce{^{15}_{\Xi^-}C}$) system named as KISO event in KEK-E373 experiment, it is found that the $\Xi^-$ binding energies are around $5.61-5.89$ MeV for $1s$ state and $0.94-1.21$ MeV for $1p$ state with QMF-NK1S, QMF-NK2S, and QMF-NK3S parameter sets, whose single $\Xi$ potentials are $-12$ MeV.  These results and those from cluster models with the Gaussian expansion method concerning on the $\ce{\Xi^{-}{+}^{12}Be}$ show that in the KISO event, the $\Xi^-$ hyperon may occupy the $1p$ state. Furthermore, the $\Xi^-$ binding energies are achieved around $27$ MeV for the $\ce{\Xi^{-}{+}^{207}Pb}$ (i.e., $\ce{^{208}_{\Xi^-}Tl}$) system. The energies were nearly comparable to the single-$\Lambda$ binding energy of $\ce{^{208}_{\Lambda}Pb}$ observed by experiments. It demonstrates that the $\Lambda$ and $\Xi^-$ hyperons seem to appear simultaneously in a neutron star.
\end{abstract}



\maketitle

\section{Introduction}

The hypernuclei and hyperon matter are increasingly greater attention in nuclear physics, both in experimental explorations and theoretical investigations. Over the past half century, there has been less experimental information concerning hypernuclei than that concerning finite nuclei due to the limitation of available technologies. Recently, many plans have been proposed to obtain more experimental data about the hypernuclei for the further researches on strangeness nuclear physics in the new-generation facilities, such as J-PARC, MAMI, JLab, and FAIR~\cite{feliciello15}. 

The $\Lambda$ hypernuclei are the most ubiquitous objects in strangeness physics with a rich experimental history, whose $\gamma$-ray spectra have been accurately measured from $\ce{_\Lambda^{3}H}$ to $\ce{_\Lambda^{208}Pb}$. So far, there is only one observed bound $\Sigma$ hypernucleus, i.e., $\ce{^4_{\Sigma}He}$ and heavier $\Sigma$ hypernuclei have not been searched out.  For an $S=-2$ system, several light double $\Lambda$ hypernuclei have been observed, such as the $\ce{^6_{\Lambda\Lambda}He},~\ce{^{10}_{\Lambda\Lambda}Be}$,  $\ce{^{12}_{\Lambda\Lambda}B}$, and $\ce{^{11}_{\Lambda\Lambda}Be}$ ($\ce{^{12}_{\Lambda\Lambda}Be}$ )~\cite{takahashi01,aoki09,ahn13}. Furthermore, there are only three events indicating the $\Xi$ hypernuclei, which are the $\ce{^{11}B} + \Xi^-$, $\ce{^{12}C} + \Xi^-$, and $\ce{^{14}N} + \Xi^-$ systems~\cite{aoki93,yamaguchi01,nakazawa15,gogami16}. The analyses of first two events were strongly dependent on the theoretical models. It is difficult to claim that there were bound $\Xi$ hypernuclei. In 2015, Nakazawa {\it et al.} confirmed a deeply bound state in the $\ce{^{14}N}+\Xi^-$ system in the KEK-E373 experiment, in which $\Xi^-+\ce{^{14}N}\rightarrow\ce{^{15}_{\Xi^-}C}\rightarrow\ce{^{10}_{\Lambda}Be}+\ce{^{5}_{\Lambda}He}$, which showed the first clear evidence of the $\Xi N$ interaction as an attractive force~\cite{nakazawa15}.  However, in Nakazawa's analysis, there were still some ambiguities to determine the $\Xi^-$ binding energy, because it could not be confirmed whether the $\ce{^{10}_{\Lambda}Be}$ was in the ground or excited states. The $\Xi^-$ binding energy would be $B_{\Xi^-}=4.38\pm0.25$ MeV if $\ce{^{15}_{\Xi^-}C}$ and $\ce{^{10}_{\Lambda}Be}$ hypernuclei are both in their ground states. However, the energy will be  $B_{\Xi^-}=1.11\pm0.25$ MeV if the  $\Lambda$ hyperon in $\ce{^{10}_{\Lambda}Be}$ occupies the excited state. In this analysis, these two binding energies were considered to correspond to two occupied states of $\Xi^-$ in the $\ce{^{14}N}+\Xi^-$ system as the $1s$ and $1p$ orbits, respectively.  Therefore, it is very important to investigate the binding energies of the  single-$\Xi^-$ hypernuclei from the aspect of theoretical methods in the likely event of insufficient experimental information.

The light $\Xi^-$ hypernuclei ($A<12$) have been described by the microscopic cluster model by using the Gaussian expansion method with the effective $\Xi$ potential obtained from the $G$-matrix of the Nijmegen baryon-baryon interaction~\cite{hiyama08}. The heavier $\Xi^-$ hypernuclei ($\Xi^-+\ce{^{12}C},~\Xi^-+\ce{^{14}N}$ and $\Xi^-+\ce{^{16}O}$ systems) were studied with the local density approximation from the $G$-matrix of the Nijmegen hard-core model D potential~\cite{yamaguchi01}. Furthermore, the heavier $\Xi^-$ hypernuclei were also systematically calculated by using the relativistic mean-field (RMF) method~\cite{mares94,tan04} and the quark-meson coupling model~\cite{tsushima98,guichon08}  from $\Xi^-+\ce{^{16}O}$ to $\Xi^-+\ce{^{208}Pb}$ systems.  In these relativistic many-body methods, the coupling constants between the $\Xi$ hyperon and mesons were determined by assuming a reasonable single-$\Xi$ potential at the nuclear saturation density with SU(3) symmetry. Therefore, the properties of the $\Xi^-$ hypernuclei in these efforts were strongly dependent on the choices of the magnitudes of the $\Xi$ potential at nuclear saturation density and were not constrained by the experimental data.

Recently, Sun {\it et al.}  adopted the RMF method and Skyrme-Hartree-Fock (SHF) model to discuss the $\Xi^-$ binding energies of  $\Xi^-+\ce{^{14}N}$ within the constraints of Nakazawa's analysis~\cite{sun16}. Sun {\it et al.} found that the $\Xi^-$ hyperon needed to occupy the $1p$ state in the KISO event to satisfy the ``empirical data" of the $\Xi^-$ binding energy in  the $\Xi^-+\ce{^{12}Be}$ system of approximately $3.0-5.5$ MeV with the same interaction. In this sense, the $\Xi$ potential in nuclear matter can be determined, which would be of great help in understanding the role of the $\Xi$ hyperon in neutron stars.

Over the past few years, we have developed a quark mean field (QMF) model to study the nuclear many-body systems from finite nuclei to neutron stars~\cite{toki98,shen00,shen02,hu14a,hu14b,xing16,xing17}. In the QMF model, the baryons are constructed of three constituent quarks with central confinement potentials. The baryons interacts with each other by exchanging the light mesons, such as $\sigma,~\omega$, and $\rho$ mesons between the quarks located in different baryons. Three nucleon-nucleon interactions for the QMF model (QMF-NK1, QMF-NK2, and QMF-NK3), which corresponded to different masses of constituent quarks,  were obtained by fitting the binding energies and charge radii of several double-magic nuclei~\cite{xing16}.  The properties of other nuclei and nuclear matter are very well described by these three parameters in the QMF model. Furthermore, the strangeness degree of freedom was also included in the QMF model to generate three baryon-baryon interactions, the  QMF-NK1S, QMF-NK2S, and QMF-NK3S~\cite{xing17}. The coupling constants between the mesons and hyperons were determined by empirical values of hyperon-nucleon potentials at nuclear saturation density, $U^{(N)}_{\Lambda}=-30$ MeV and  $U^{(N)}_{\Xi}=-12$ MeV. 
It was found that the binding energies of the single-$\Lambda$ hypernuclei obtained from experimental data could be reproduced very well by the QMF model  and the maximum mass of  a neutron star containing the hyperons was approximately $2.1M_\odot$ as derived from these three interactions. These results concerning neutron stars agreed with the requirements of recent astronomical observations concerning the massive neutron stars~\cite{demorest10,antoniadis13,fonseca16}. 

The properties of several heavy $\Xi^0$ hypernuclei were also investigated in our previous work, where the $\Xi^0\rho$ coupling was neglected for simplicity. The deepest bound state was  the $1s$ state of $\ce{^{208}_{\Xi^0}Pb} $, whose single-$\Xi^0$ binding energy was approximately $9.5$ MeV. In this current work, we sought to apply the QMF model to study the properties of $\Xi^-$ hypernuclei from light to heavy mass regions within QMF-NK1S, QMF-NK2S, and QMF-NK3S interactions. The $\Xi^-\rho$ coupling will be included which is essential for the charged $\Xi^-$ hyperon. Because of its negative charge, it is expected that the $\Xi^-$ hypernuclei will be bound more tightly than $\Xi^0$ hypernuclei. The $\Xi^-+\ce{^{14}N}$ system will be discussed especially to  confirm the occupied state of $\Xi^-$ hyperon in KISO event.

This paper is arranged as follows. In Sec. II, we briefly show the formulas of the QMF model for the $\Xi^-$ hypernuclei. In Sec. III, the properties of $\Xi^-$ hypernuclei will be calculated systematically. The KISO event is also discussed in detail. Finally, a summary is given in Sec. IV.

\section{Quark mean field model for $\Xi^-$ hypernuclei}
In the QMF model, the baryons are composed of three constituent quarks. Accordingly, the $\Xi^-$ hyperon is combined with one $d$ quark and two $s$ quarks, which are confined via central confinement potentials. Due to the non-perturbative character of the QCD theory in the low energy region, the details about the confinement potentials are still not well understood. In this work, a harmonic oscillator form that considers the scalar-vector Dirac structure, $U(r)=\frac{1}{2}(1+\gamma^0)(a_qr^2+V_q)$, is used. In this instance, the wave function of the quark is solved analytically. In principle, the masses of baryons will be generated by the quark energies and the  center of mass corrections. To treat the effects of chiral symmetry and gluon exchange between quarks, the pionic and gluonic corrections are also included in the baryon masses by using perturbation theories. The details of how to construct the baryon masses from quark level can be found in our previous works~\cite{xing16,xing17}. 

Based to the classical scheme of one-boson-exchange potential, the baryons in the QMF model interact with each other by exchanging the light mesons, such as $\sigma,~\omega$ and $\rho$, between quarks in distinguished baryons. Therefore, at a finite density, the scalar meson, $\sigma$ will be embraced in the quark mass that wll be presented as the effective quark mass. The baryon masses will be strongly related to the strength of $\sigma$ meson at the hadron level as a function, $M^*_B(\sigma)$. This is very similar to the mechanism of the effective baryon masses in the RMF model, which are defined as $M^*_B(\sigma)=M_B+g_{\sigma B}\sigma$.

Consequently, the Lagrangian of the QMF model for $\Xi^-$ hypernuclei is written as an analogous form in the RMF model that is associated with the nucleon ($\psi_N$), $\Xi$ hyperon ($\Xi$), scalar-isoscalar meson ($\sigma$), vector-isoscalar meson ($\omega$), vector-isovector meson ($\rho$) and the photon ($A$) fields as expressed by the following~\cite{toki98,shen00,shen02,hu14a,hu14b,xing17},
\beq
{\cal L}_{\rm QMF}
&=&
\bar\psi_N\left[ i\gamma_\mu\partial^\mu-M_N^*
-g_{\omega N}\gamma^\mu \omega_\mu 
-g_{\rho N} \gamma^\mu\vec\tau_N\cdot\vec\rho_\mu 
-e\frac{(1+\tau_{N,3})}{2}\gamma^\mu A_\mu 
\right]\psi_N\nn
&&
+\bar\psi_\Xi
\left[i\gamma_\mu\partial^\mu-M_\Xi^*
-g_{\omega\Xi}\omega_\mu\gamma^\mu
+\frac{f_{\omega\Xi}}{2M_\Xi}\sigma^{\mu\nu}\partial_\nu\omega_\mu
-g_{\rho \Xi} \gamma^\mu\vec\tau_\Xi\cdot\vec\rho_\mu 
-e\frac{(\tau_{\Xi,3}-1)}{2}\gamma^\mu A_\mu \right]\psi_\Xi\nn
&&
+\frac{1}{2} \partial^\mu\sigma\partial_\mu\sigma
-\frac{1}{2} m_\sigma^2\sigma^2
-\frac{1}{3} g_2\sigma^3
-\frac{1}{4} g_3\sigma^4\nn
&&
-\frac{1}{4} W^{\mu\nu}W_{\mu\nu}
+\frac{1}{2} m_\omega^2\omega^2
+\frac{1}{4} c_3\omega^4\nn
&&
-\frac{1}{4} \vec R^{\mu\nu}\vec R_{\mu\nu}
+\frac{1}{2} m_\rho^2\rho^2
-\frac{1}{4}F^{\mu\nu}F_{\mu\nu},
\eeq
where the arrows denote the isospin vectors and three tensor operators for the vector and the photon fields are defined as follows,
\beq
W^{\mu\nu}&=&\partial^\mu\omega^\nu-\partial^\nu\omega^\mu,\nn
\vec R^{\mu\nu}&=&\partial^\mu\vec\rho^\nu-\partial^\nu\vec\rho^\mu,\nn
F^{\mu\nu}&=&\partial^\mu A^\nu-\partial^\nu A^\mu.
\eeq 
The effective masses of the nucleon and $\Xi$ hyperon, $M^*_N$ and $M^*_\Xi$, respectively, are created by the constituent quark model with confinement potentials. The tensor coupling between the $\omega$ meson and the $\Xi$ hyperon, $\frac{f_{\omega\Xi}}{2M_\Xi}\sigma^{\mu\nu}\partial_\nu\omega_\omega$, is considered to improve the description of the small spin-orbit splittings of the hypernuclei in the experimental data~\cite{sugahara94,shen06}. $\tau_{N,3}$ is the third component of nucleon isospin operator, $\vec\tau$. Moreover $\tau_{N,3}=1$ for proton and $\tau_{N,3}=-1$ for neutron in conventional calculations, while for the isospin operator of the $\Xi$ hyperon, $\tau_{\Xi,3}=\pm1$ for $\Xi^0$ and $\Xi^-$, respectively. 

In this work, the $\Xi^-$ hypernuclei are treated as spherical cases approximately and the spatial components of vector meson vanish in the time-reversal symmetry. There are only time components of the $\omega,~\rho$ and $A$ fields in Lagrangian. For the convenient presentation later on, we will use the symbols, $\omega,~\rho$, and $A$,  instead of $\omega^0,~\rho^0$, and $A^0$.

The equations of motion about the nucleon, $\Xi$ hyperon and mesons can be obtained within the Euler-Lagrange equations. However, these equations of motion for quantum fields cannot be exactly solved. The mean-field approximation and no-sea approximation will be employed in considering the mesons as classical fields in QMF model. Then, the Dirac equations for the nucleons and the $\Xi$ hyperon can be expressed as below,
\beq
&&\left[i\gamma_{\mu}\partial^{\mu}-M_N^*
-g_{\omega N}\omega\gamma^0
-g_{\rho N}\rho\tau_{N,3}\gamma^0\
-e\frac{(1+\tau_{N,3})}{2}A\gamma^0
\right]\psi
=0,\nn
&&\left[i\gamma_{\mu}\partial^{\mu}-M_\Xi^*
-g_{\omega\Xi}\omega\gamma^0
+\frac{f_{\omega\Xi}}{2M_\Xi}\sigma^{0i}\partial_i\omega
-g_{\rho \Xi}\rho\tau_{\Xi,3}\gamma^0-e\frac{(\tau_{\Xi,3}-1)}{2}A\gamma^0\right]\psi_\Xi
=0
\eeq
and the equations of motion for mesons are given by
\beq
&&-\Delta\sigma+m_\sigma^2\sigma+g_2\sigma^2+g_3\sigma^3
=-\frac{\partial M_N^*}{\partial\sigma}
\langle\bar\psi_N\psi_N\rangle
-\frac{\partial M_\Xi^*}{\partial\sigma}
\langle\bar\psi_\Xi\psi_\Xi\rangle,\nn
&&-\Delta\omega+m_\omega^2\omega+c_3 \omega^3=
g_{\omega N}\langle\bar\psi_N\gamma^0\psi_N\rangle
+g_{\omega\Xi}
\langle\bar\psi_\Xi\gamma^0\psi_\Xi\rangle
-\frac{f_{\omega\Xi}}{2M_\Xi}
\partial_i\langle\bar\psi_\Xi\sigma^{0i}
\psi_\Xi\rangle,\nn
&&-\Delta\rho+m_\rho^2\rho=
g_{\rho N}\langle\bar\psi_N\tau_{N,3}\gamma^0\psi_N\rangle
+g_{\rho \Xi}\langle\bar\psi_\Xi\tau_{\Xi,3}\gamma^0\psi_\Xi\rangle,\nn
&&-\Delta A=
e\langle\bar\psi_N\frac{(1+\tau_{N,3})}{2}\gamma^0\psi_N\rangle+e\langle\bar\psi_\Xi\frac{(\tau_{\Xi,3}-1)}{2}\gamma^0\psi_\Xi\rangle.
\eeq
These coupling equations are solved self-consistently for $\Xi^-$ hypernuclei with numerical methods, when a hypernucleus is regarded as a core of finite nuclei plusing one $\Xi^-$ hyperon. If the core is an open shell nuclei, the pairing effect will be taken into account by employing BCS theory. The center of mass correction for hypernucleus in this work is dealt with the microscopic method as~\cite{shen06},  
\beq
E_{\text{c.m.}}\frac{\langle \Psi|\vec P^2_{\text{c.m.}}|\Psi\rangle}{2M_{\text{total}}},
\eeq
where, $\Psi$ is the total wave function of the entire system. $M_{\text{total}}$ is the total mass of the hypernucleus and $\vec P_{\text{c.m.}}$ is the total momentum operator.
\section{Results and discussion}
In the QMF model, the coupling strengths between the $\sigma$ meson and the baryons are determined by the effective baryon masses generated from the confinement potentials of the three quarks. Three constituent quark masses ($m_q=250, ~300$, and $350$ MeV) were chosen to consider the quark mass dependence of the baryons. Three corresponding parameter sets (QMF-NK1, QMF-NK2, and QMF-NK3) at the hadron levels that were related to the coupling constants between the vector mesons and baryons, were obtained by fitting the ground-state properties of several double magic nuclei, i.e., $\ce{^{40}Ca},~\ce{^{48}Ca},~\ce{^{90} Zr}$, and $\ce{^{208}Pb}$~\cite{xing16}.  These parameter sets produced excellent descriptions of the finite nuclei and nuclear matter. 

The strangeness degree of freedom was then included to study the single-$\Lambda,~\Xi^0$ hypernuclei, and the neutron star with hyperons~\cite{xing17}. The coupling constants between the $\omega$ meson and the $\Lambda,~\Xi$ baryons were provided by the empirical values of the single-$\Lambda$ and $\Xi$ potentials at the nuclear saturation density due to the insufficiency of the experimental data for the hypernuclei. The single-$\Lambda$ potential was considered to be $U^{(N)}_{\Lambda}=-30.0$ MeV, which reproduced the experimental data concerning the binding energies of the single-$\Lambda$ hypernuclei very well. For the $\Xi$ system, the  $U^{(N)}_{\Xi}=-12.0$ MeV was used at the nuclear saturation density. Finally, these coupling constants were termed QMF-NK1S, QMF-NK2S, and QMF-NK3S parameter sets.  In these cases, the maximum masses of the neutron stars with hyperons approached to $2.1 M_\odot$, which were satisfied the constraints of recent astronomical observations in massive neutron stars~\cite{demorest10,antoniadis13,fonseca16}.  Furthermore, it was found that the $\Lambda$ and $\Xi^-$ hyperons simultaneously appeared in the core region of  the neutron stars.

In the present work, we concentrate on studying the properties of the $\Xi^-$ hypernuclei, in particular about their $\Xi^-$ binding energies. To discuss the influence of different $\Xi N$ interactions on the $\Xi^-$ hypernuclei, an additional three coupling constants between the $\omega$ meson and the $\Xi$ hyperon were fixed to generate the $U^{(N)}_{\Xi}=-9.0$ MeV at saturation density, which were termed as QMF-NK1S', QMF-NK2S', and QMF-NK3S', respectively. The $\rho$ meson should be considered due to the isospin character of the $\Xi$ hyperon. However, the coupling constant to the $\Xi$ hyperon cannot be easily determined by the single $\Xi$ potential at nuclear matter. The SU(3) symmetry is employed to generate, $g_{\rho\Xi}=g_{\omega\Xi}$.  In addition, the spin-orbit splittings of hypernuclei were found to be smaller than the ones in finite nuclei. The tensor coupling terms between the $\omega$ and the $\Xi$ hyperon are considered as, $f_{\omega\Xi}=-0.4g_{\omega\Xi}$~\cite{mares94,sun16}.  

The contribution of the $\rho$ meson in the single-$\Xi^-$ hypernuclei is treated carefully, in particular, for the pure isospin-zero core, where the $\rho$ meson is merely brought by the $\Xi^-$ hyperon. However, there is only one $\Xi^-$ hyperon in a single-$\Xi^-$ hypernucleus, where the effect of the $\rho$ meson is spurious in the Hartree approximation and is removed. Following the scheme in Refs. ~\cite{mares94,sun16}, this spurious contribution is obtained by comparing the calculations at $g_{\rho N}=0$ and  $g_{\rho N}=g_{\rho \Xi}=0$. The final result will be produced by subtracting the spurious contributions in the full self-consistent calculations.

\begin{figure}[htb]
	\centering
	\includegraphics[width=12cm]{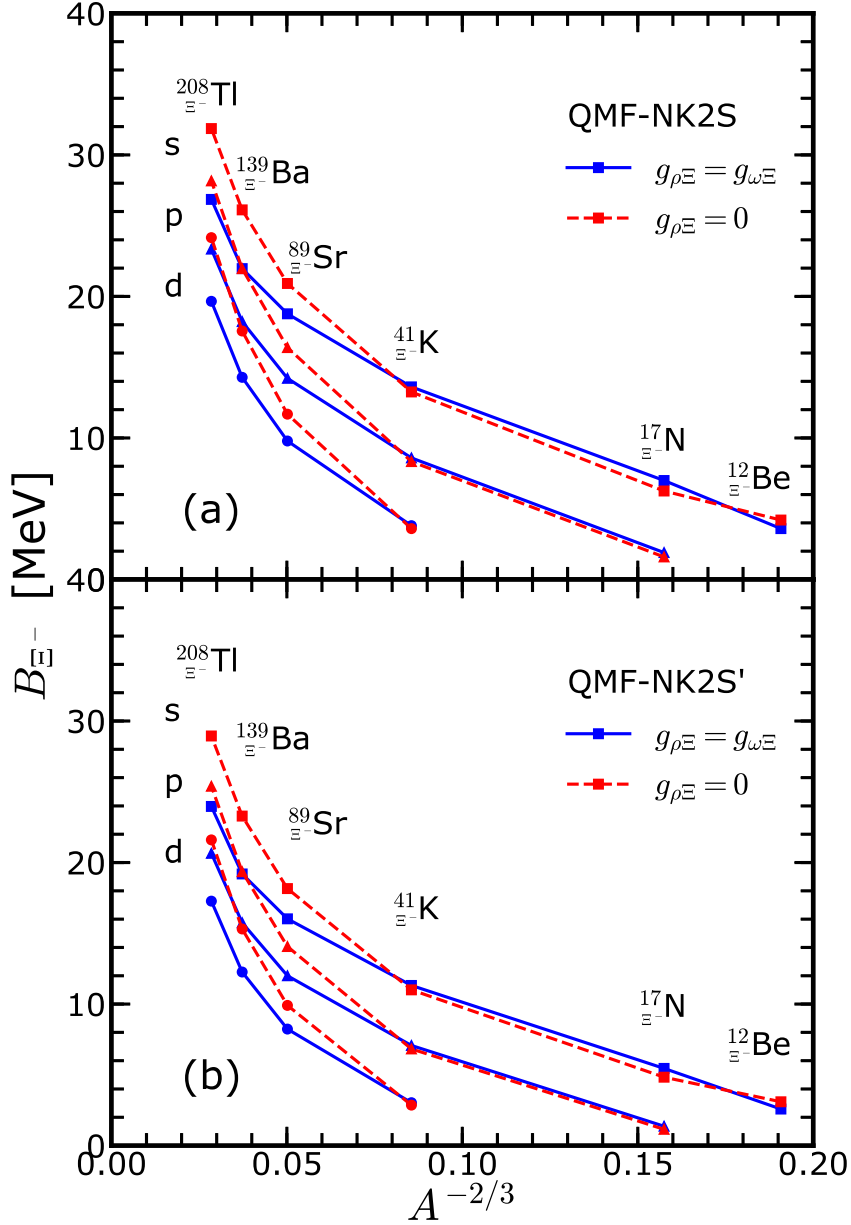}
	\caption{(Color online) Systematic calculations of the binding energies of  $\Xi^-$ hypernuclei within QMF-NK2S ( panel (a)) and QMF-NK2S' (panel (b)) parameter sets from $\ce{_{\Xi^-}^{12}Be}$ to $\ce{_{\Xi^-}^{208}Tl}$.}
	\label{xie}
\end{figure}

In Fig.~\ref{xie}, the $\Xi^-$ binding energies of a single-$\Xi^-$ hypernuclei are calculated systematically from $\ce{_{\Xi^-}^{12}Be}$ to $\ce{_{\Xi^-}^{208}Tl}$ at $s, ~p$, and $d$ orbits with QMF-NK2S (panel (a)) and QMF-NK2S' (panel (b)) parameter sets. The results that are obtained from another interactions are found to be very similar. Since the $\Xi^-$ hyperon contains one negative charge, the actual chemical symbols for the $\Xi^-$ hypernuclei do not correspond to those for the elements of the core. For instance, the $\ce{_{\Xi^-}^{208}Tl}$ denotes the $\Xi^-+\ce{^{207}Pb}$ system. To show the role of the $\rho$ meson in the $\Xi^-$ hypernuclei more clearly, results without the $\rho\Xi$ coupling case $g_{\rho\Xi}=0$ are given as dashed lines to compare with the results obtained by a proper treatment, which are plotted as solid curves.

The attractive contribution of the Coulomb field in the $\Xi^-$ hyperon causes the binding energies of the $\Xi^-$ hypernuclei to be much larger than those of the $\Xi^0$ hypernuclei. In our previous work, it was demonstrated that the $\Xi^0$ binding energy was approximately $9.5$ MeV for $\ce{_{\Xi^0}^{208}Pb}$ at the $1s$ state~\cite{xing17}. However, it is $31.86$ MeV for $\ce{_{\Xi^-}^{208}Tl}$ with a QMF-NK2S parameter set with $g_{\rho\Xi}=0$. When the $\rho$ meson is considered, the single-$\Xi^-$ binding energy of the $\ce{_{\Xi^-}^{208}Tl}$ is $26.86$ MeV at the $1s$ state using the QMF-NK2S parameter set. This is comparable to the single-$\Lambda$ binding energy of the $\ce{_{\Lambda}^{208}Pb}$, i.e., $25.95$ MeV. This result suggests that the $\Lambda$ and $\Xi^-$ hyperons may simultaneously appear in the core region of neutron stars. This conclusion was confirmed in our previous calculations concerning neutron star with hyperons. In the high density regions of a neutron stars, the fraction of the $\Xi^-$ hyperon is even larger than that of the $\Lambda$ hyperon. With the reduction of the $\Xi^-$ hypernuclei masses, the binding energies of the $\Xi^-$ hyperon rapidly decrease. This energy drops off to approximately $4$ MeV for the $\ce{_{\Xi^-}^{12}Be}$.

The contribution of the $\rho$ meson is repulsive for the $\Xi^-$ hypernuclei with a neutron-rich core, which is obtained by comparing the binding energies of the $\Xi^-$ hyperons in the $g_{\rho\Xi}=0$ and $g_{\rho\Xi}=g_{\omega\Xi}$ cases. This repulsive effect is mainly generated by the difference between neutrons and protons. This energy is shifted by $5.00$ MeV for $\ce{_{\Xi^-}^{208}Tl}$ in the QMF-NK2S. For those $\Xi^-$ hypernuclei with $Z=N$ cores ($\ce{_{\Xi^-}^{41}K}$ and $\ce{_{\Xi^-}^{17}N}$), the $\rho$ meson plays a negligible role, where the binding energies of $\Xi^-$ hyperons in $g_{\rho\Xi}=0$ and $g_{\rho\Xi}=g_{\omega\Xi}$ cases are almost identical after the spurious $\rho$-coupling effect is removed. In the panel (b) of Fig.~\ref{xie}, the results are shown for the QMF-NK2S' interaction, where the $U^{(N)}_{\Xi}=-9.0$ at nuclear saturation density is chosen. The binding energy of the $\ce{_{\Xi^-}^{208}Tl}$ at the $1s$ state is reduced by approximately $3.0$ MeV compared to that of the QMF-NK2S, which is consistent with the decrement of the single $\Xi$ potential at nuclear saturation density. For the light $\Xi^-$ hypernuclei, these reduction are smaller.

Currently, only three events were observed for the $\Xi^-$ hypernuclei in accelerators. Recently, the $\ce{_{\Xi^-}^{15}C}$ were confirmed by Nakazawa {\it et al.} as KISO event~\cite{nakazawa15}. However, it is very difficult to determine the accuracy of the binding energy due to the unknown details of $\ce{_{\Lambda}^{10}Be}$, which is produced from the decay of $\ce{_{\Xi^-}^{15}C}$. Therefore, we attempt to discuss the $\ce{_{\Xi^-}^{15}C}$ in detail in the present framework to benefit the analysis of the KISO event.

The binding energies of the $\Xi^-$ hyperon of the $\ce{_{\Xi^-}^{15}C}$ and $\ce{^{12}_{\Xi^-}Be}$ are listed on the basis of six parameter sets as previously mentioned in Table~\ref{tab} and compared to the probable experimental data. In the QMF-NK1S' to QMF-NK3S' interactions, the $B_{\Xi^-}$ of the $1s$ state is located at $4.24-4.45$ MeV for $\ce{_{\Xi^-}^{15}C}$ when all the mesons are considered. These will decrease slightly in the case of the removal of the $\rho$ meson as $g_{\rho\Xi}=0$. The $\rho$ meson contribution is very weak in $\ce{_{\Xi^-}^{15}C}$ because its core was composed of $Z=N=7$. The similar situation is arisen in the $1p$ state, whose binding energies are almost the same with and without $\rho$ meson. They are between $0.75-0.84$ in the QMF-NK1S' to QMF-NK3S' sets. With the same $\Xi N$ interaction, it is found that the $B_{\Xi^-} (1s)$ of the $\ce{^{12}_{\Xi^-}Be}$ are around $2.49-2.75$ MeV.

When the $U^{(N)}_{\Xi}$ is changed as $-12$ MeV at nuclear saturation density, i.e., QMF-NK1S, QMF-NK2S, and QMF-NK3S sets used, $B_{\Xi^-} (1s)$ of $\ce{_{\Xi^-}^{15}C}$ are between $5.61-5.82$ MeV when the $\rho$ meson effect is included, while they become as $5.58-5.80$ MeV without the couplings between $\rho$ meson and $\Xi^-$ hyperon. The corresponding $B_{\Xi^-} (1p)$ are between $1.08-1.21$ MeV in QMF-NK1S to QMF-NK3S sets, respectively. Now, the $B_{\Xi^-} (1s)$ of the $\ce{^{12}_{\Xi^-}Be}$ were calculated to be $3.49-3.78$ MeV. Furthermore, the results for switching off the $\Xi N$ Coulomb interaction also helped fill in the parentheses of Table~\ref{tab}.

In addition, in the analysis of Nakazawa {\it et al.} of the KISO event~\cite{nakazawa15}, if the $\Xi^-$ hyperon occupies $1s$ state, then its binding energy is $4.38\pm0.25$ MeV. Moreover, the value will become $1.11\pm0.25$ MeV at the $1p$ state. Therefore, it is difficult to directly distinguish the angular momentum of the $\Xi^-$ hyperon of the KISO event based entirely on the theoretical results of the $\ce{_{\Xi^-}^{15}C}$ in the QMF model. However, there is also some evidence of $\ce{^{12}_{\Xi^-}Be}$ in the experiments. Furthermore, the binding energy of $\ce{^{12}_{\Xi^-}Be}$ was approximately $5$ MeV with Coulomb interaction and $2.2$ MeV without in a cluster-model calculation~\cite{hiyama08}. Moreover, the energy was approximately $3.0-5.5$ MeV with an antisymmetrized molecular dynamics approach~\cite{matsumiya11}. In present calculation, they are about $3.49-3.78$ MeV and $4.11-4.35$ MeV, corresponding to be with and without $\rho$ meson contributions, respectively in QMF-NK1S to QMF-NK3S interactions. Therefore, the $\ce{_{\Xi^-}^{15}C}$ in this event is more likely to be a $\ce{^{14}N}+\Xi^-(1p)$ system in terms of the predictions of our theoretical framework, which is consistent with the conclusion reached by Sun {\it et al.} using the RMF model~\cite{sun16}, where the coupling constant between $\sigma$ meson and $\Xi$ hyperon was adjusted to reproduce the $\Xi$  binding energy and the one between $\omega$ and $\Xi$  was decided by the quark counting rules, while the coupling between $\sigma$ and $\Xi$ in the QMF model is generated by the constituent quark model and is varied with density. 

Note that in the RMF calculation, when the $\Xi^-$ hyperon binding energy of the $1p$ state is found to be $1.1$ MeV, the corresponding value of the $1s$ state is approximately $9.4$ MeV. This difference is much larger than the case in the QMF model, where $B_{\Xi^-}(1s)-B_{\Xi^-}(1p)$ is just approximately $4.6$ MeV. Furthermore, the $B_{\Xi^-}(1s)$ in QMF model is only $30\%$ greater than the experimental data. Actually, these results are strongly dependent on the various mean fields obtained from mesons and photons, which will be discussed later. Through this analysis, it was demonstrated that $U^{(N)}_{\Xi}$ is  $12$ MeV at nuclear saturation density that is quite reasonable.

\begin{table}[htb]
	\centering
	\caption{The binding energies (in MeV) of $\ce{^{12}_{\Xi^-}Be}$ and $\ce{_{\Xi^-}^{15}C}$ for different orbital states of the $\Xi^-$ hyperons in different QMF interactions and the corresponding experimental data. The values in parentheses are calculated in the case of removing the Coulomb interaction between the $\Xi^-$ hyperon and nucleon.}
	\label{tab}
	\begin{tabular}{l c c c c}
		\hline
		\hline
		                   &~~~~~$\ce{^{15}_{\Xi^-}C}(1s)$~~~~~&~~~~~$\ce{^{15}_{\Xi^-}C}(1p)$~~~~~&~~~~~$\ce{^{12}_{\Xi^-}Be}(1s)$~~~~~\\
		\hline		
		QMF-NK1S($g_{\rho\Xi}=g_{\rho N}$)&5.82 (2.79) &1.21 (-0.56)      &3.78 (1.77)    \\
		QMF-NK2S($g_{\rho\Xi}=g_{\rho N}$)&5.69 (2.66) &1.14 (-0.57)     &3.59 (1.60)    \\
		QMF-NK3S($g_{\rho\Xi}=g_{\rho N}$)&5.61 (2.58) &1.08 (-0.59)    &3.49 (1.49)    \\
		\hline	
		QMF-NK1S($g_{\rho\Xi}=0$)&5.80 (2.80)   &1.21  (-0.56)     &4.35 (2.27)    \\
		QMF-NK2S($g_{\rho\Xi}=0$)&5.65 (2.66)  &1.14  (-0.57)     & 4.20 (2.13)   \\
		QMF-NK3S($g_{\rho\Xi}=0$)&5.58 (2.58)  &1.08 (-0.59)    & 4.11 (2.03)    \\	
		\hline			
		QMF-NK1S'($g_{\rho\Xi}=g_{\rho N}$)&4.45 (1.62) &0.84 (-0.63)    &2.75 (0.91)   \\
		QMF-NK2S'($g_{\rho\Xi}=g_{\rho N}$)&4.31 (1.50) &0.79  (-0.64)   &2.58  (0.77)  \\
		QMF-NK3S'($g_{\rho\Xi}=g_{\rho N}$)&4.24 (1.43)&0.75  (-0.64)   &2.49  (0.70)  \\
		\hline	
		QMF-NK1S'($g_{\rho\Xi}=0$)                &4.44  (1.64)&0.84 (-0.63)     &3.23  (1.30)  \\
		QMF-NK2S'($g_{\rho\Xi}=0$)               &4.30  (1.52)&0.79  (-0.64)    &3.10   (1.18) \\
		QMF-NK3S'($g_{\rho\Xi}=0$)               &4.24  (1.44)&0.75  (-0.64)   &3.01    (1.09)\\	
		\hline					
		Expt. or empirical data & $4.38\pm0.25$~\cite{nakazawa15}& $1.11\pm0.25$~\cite{nakazawa15}&$3.0-5.5$~\cite{matsumiya11}\\
		\hline
		\hline
	\end{tabular}
\end{table}

\begin{figure}[htb]
	\centering
	\includegraphics[width=12cm]{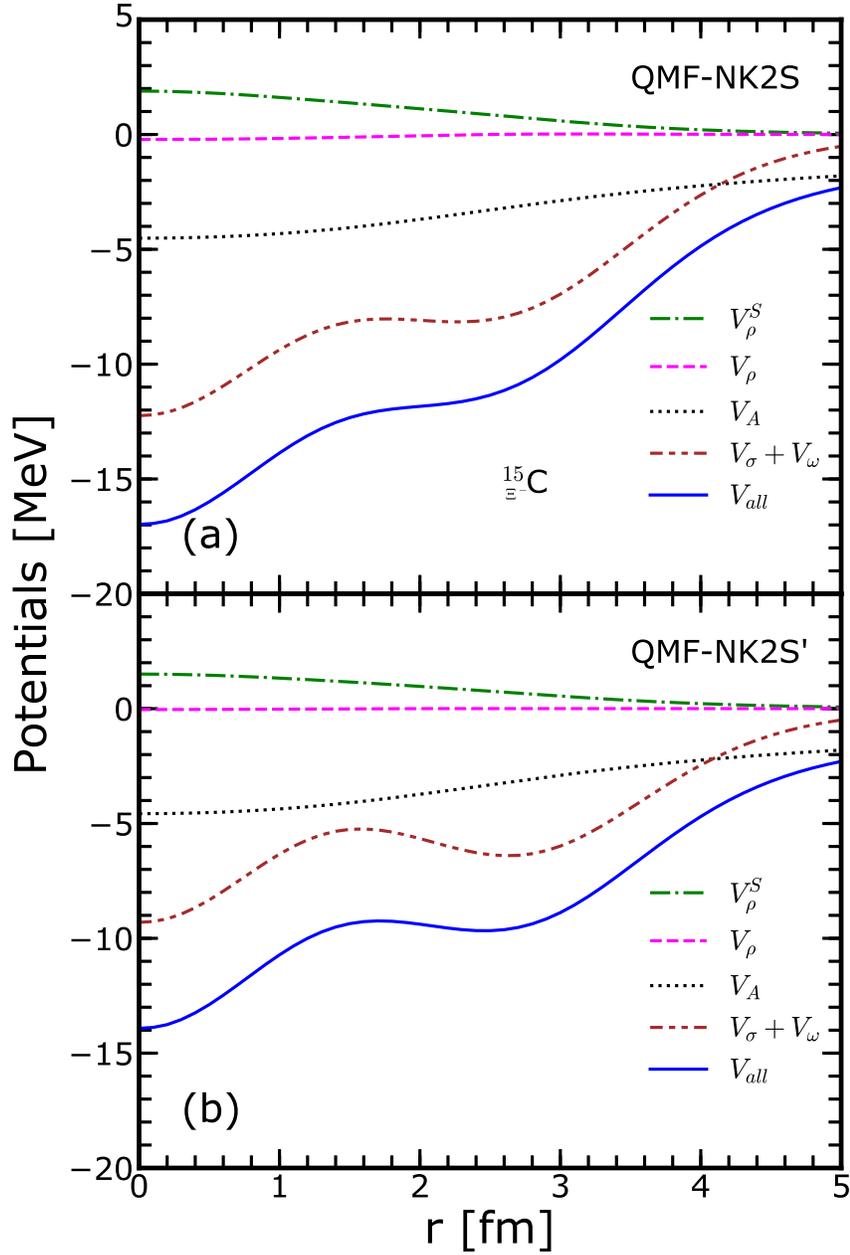}
	\caption{(Color online) The potentials from various fields in $\ce{^{15}_{\Xi^-}C}$ within QMF-NK2S and QMF-NK2S'. The dot-dashed ones are from the spurious contributions of $\rho$ meson and dashed curves for its real effects. Dotted lines represent Coulomb field. The dash-dot-dotted curves denote the potentials from $\sigma$ and $\omega$ mesons. The potentials from all mesons are given as solid curves, where $V_{all}=V_\sigma+V_\omega+V_\rho+V_A$.}
	\label{pot}
\end{figure}

In Fig.~\ref{pot}, the various potentials from different meson fields in $\ce{^{15}_{\Xi^-}C}$ are shown with QMF-NK2S (panel (a)) and QMF-NK2S' (panel (b)) interactions. The total contributions of the $\sigma$ and $\omega$ mesons provide the main feature. In the center region of $\Xi^-$ hypernucleus, the magnitudes of $V_\sigma+V_\omega$ is approximately $-12.0$ MeV in the QMF-NK2S interaction, which represents the single-$\Xi$ potential of infinite nuclear matter at the nuclear saturation density. The negative charge of the $\Xi^-$ hyperon generated a $-5.0$ MeV attractive contribution for the Coulomb interaction. This is a long range force, which converges slowly. The spurious components of $\rho$ meson are denoted as $V_\rho^S$ in the full calculations, which consider a repulsive effect of approximately $2.0$ MeV at the center region. The real $\rho$ meson potential is obtained after removing the $V_\rho^S$ part, and it becomes slightly attractive. In total, in the center region, there was a mean-field potential of approximately $-17.0$ MeV in QMF-NK2S.

To compare the potentials from RMF model in the work of Sun {\it et al.}, the mean-field potentials from QMF-NK2S' are shown in the panel (b), which also generated the $B_{\Xi^-}(1s)$ of $\ce{^{12}_{\Xi^-}Be}$ at approximately $4.4$ MeV. The obvious difference between the QMF and RMF models appears in the intermediate region ($1-3$ fm) in the potentials for the $\sigma$ and $\omega$ mesons. In the center part, the magnitudes of $V_\sigma+V_\omega$ is around $-9.0$ MeV. Then, the magnitude increases to $-5.0$ MeV at $ r=1.5$ fm in the QMF model and decreases slowly between $1.5$ fm and $3.0$ fm, while the deepest point of $V_\sigma+V_\omega$ is around $-12.0$ MeV and eases up at $2.0$ fm in RMF model~\cite{sun16}. The mean fields in QMF model are shallower but wider, while they are deeper but narrower in RMF model. This is the reason why the two different models can provide similar binding energies for the $1s$ states, while those of $1p$ states are completely different. 

\begin{figure}[htb]
	\centering
	\includegraphics[width=12cm]{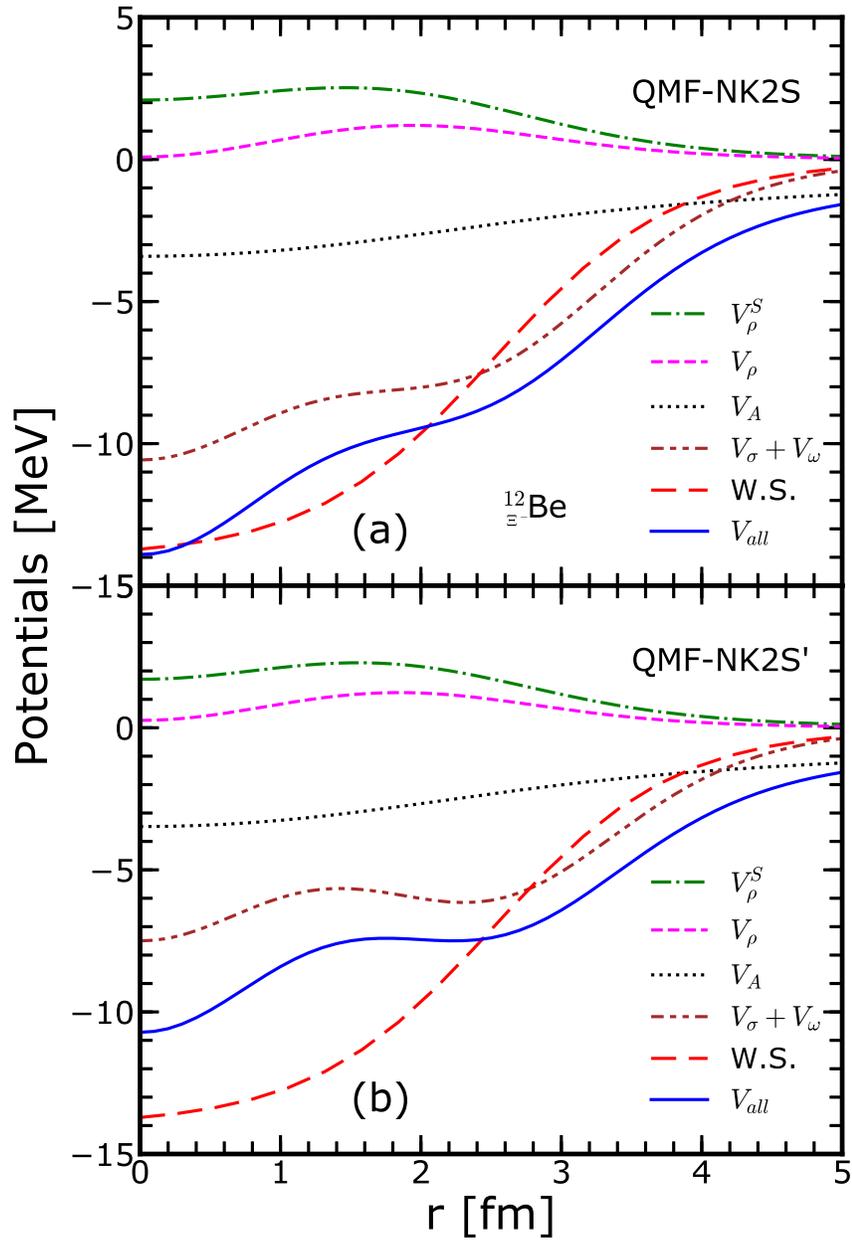}
	\caption{(Color online) The potentials from various fields in $\ce{^{12}_{\Xi^-}Be}$ within QMF-NK2S and QMF-NK2S'. The WS potential used in the analysis of Nakazawa {\it et al.}~\cite{nakazawa15} is also plotted for comparison as long dashed curves. }
	\label{potb}
\end{figure}

In Fig.~\ref{potb}, the mean-field potentials for $\ce{^{12}_{\Xi^-}Be}$ are plotted within the QMF-NK2S (panel (a)) and QMF-NK2S' (panel (b)) parameter sets. The Woods-Saxon (WS) potential adopted in the analysis by Nakazawa {\it et al} is also given for comparison. It can be seen that the contribution of the $\rho$ meson in $\ce{^{12}_{\Xi^-}Be}$ is larger than that in $\ce{^{15}_{\Xi^-}C}$, since the core of $\ce{^{15}_{\Xi^-}C}$ ($N=Z=7$) is almost pure isospin zero. The total mean-field potential from QMF-NK2S has a similar depth as the one from the empirical WS potential.
Therefore, we obtain the $B_{\Xi^-}(1s)$ of $\ce{^{12}_{\Xi^-}Be}$ as being $3.59$ MeV, which is consistent with the result from WS potential. Conversely, the fields from QMF-NK2S' are smaller than those from QMF-NK2S and WS potential. The shape of the mean-field potential from QMF model is also different from the one obtained from the RMF theory as shown in Ref.~\cite{sun16}. At long range distance, our potentials are deeper than the WS potential. Actually, this mean-field potential is very close to that from the SL3 interaction as calculated using the SHF model in Ref.~\cite{sun16}, which is shallower at short distance but deeper at long range compared to the WS potential. The $B_{\Xi^-}(1s)-B_{\Xi^-}(1p)$ from the SL3 interaction is also smaller than the results from other interactions in Ref.~\cite{sun16}, which is in accordance with our conclusions.

\section{Conclusion}
The properties of the $\Xi^-$ hypernuclei were investigated in the framework of QMF model, where the baryons were regarded as a combination of three constituent quarks. The baryons are confined by the central harmonics oscillator potentials. At the hadron level, the baryons interact with each other by exchanging various mesons between quarks in the different baryons. In this work, two types of $\Xi N$ interactions were chosen $U^{(N)}_\Xi=-12$ MeV and $U^{(N)}_\Xi=-9$ MeV. The available $NN$ interactions, QMF-NK1, QMF-NK2, and QMF-NK3, which corresponded to three types of constituent quark masses, $m_q=250$ MeV, $m_q=300$ MeV, and $m_q=350$ MeV, were combined.
Finally, six parameter sets were obtained to study the properties of the single-$\Xi^-$ hypernuclei, QMF-NK1, QMF-NK2, QMF-NK3 ($U^{(N)}_\Xi=-12$ MeV), QMF-NK1', QMF-NK2', and QMF-NK3' ($U^{(N)}_\Xi=-9$ MeV).

The binding energies of the single-$\Xi^-$ hypernuclei were systematically studied from $\ce{_{\Xi^-}^{12}Be}$ to $\ce{_{\Xi^-}^{208}Tl}$ within the QMF-NK2S and QMF-NK2S' interactions. Since there was only one $\Xi^-$ hyperon remaining in the single-$\Xi^-$ hypernuclei, the coupling effects between the $\rho$ meson and $\Xi^-$ hyperon were properly addressed. The spurious contributions of the $\rho$ field from the Hartree approximation were removed from the conventional full calculations. This reasonable treatment found that the $\rho$ meson contribution was generated by the isospin asymmetry at neutron-rich nuclei, while its effect nearly disappeared for the $\Xi^-$ hypernuclei with a $Z=N$ core. The $B_{\Xi^-}$ of $\ce{_{\Xi^-}^{208}Tl}$ at $1s$ state were around $27$ MeV in the QMF-NK2S and $24$ MeV in QMF-NK2S' for a strong attractive force due to the Coulomb force, which was analogous to the $B_\Lambda$ of $\ce{_{\Lambda}^{208}Pb}$. Therefore, in the core region of neutron stars, the $\Lambda$ and $\Xi^-$ hyperons will appear to have almost the same density. Furthermore, these binding energies were found to be weakly dependent on the QMF interactions. 

The KISO event as a bound state of the $\ce{^{14}N}+\Xi^-$ system was also discussed in detail. It was found that the results of the QMF-NK2S interaction supported that the $\Xi^-$ hyperon should occupy the $1p$ states in this event. This result is in accordance with the recent conclusions by using the RMF model, where the Kiso event was regarded as an observation of the excited state in $\ce{^{15}_{\Xi}C}$~\cite{sun16}. However, in the QMF model the $B_{\Xi^-}(1s)-B_{\Xi^-}(1p)$ of $\ce{^{15}_{\Xi^-}C}$ were much smaller than those in RMF model, which were generated by the different behaviors of $\sigma$ and $\omega$ potentials in these two models. In the present framework, the binding energy of another event about $\Xi^-$ hypernuclei, $\ce{_{\Xi^-}^{12}Be}$, was similar to the calculations from cluster models with the Gaussian expansion method and antisymmetrized molecular dynamics approach.

The description of light $\Xi^-$ hypernuclei within the present QMF model was in good agreement with the RMF model, cluster model, and antisymmetrized molecular dynamics model . For the heavy $\Xi^-$ hypernuclei, it was found that they were located in very deep bound states. Therefore, in a large mass region, the $\Xi^-$ hypernuclei may be measured easily in the facility.

\section*{Acknowledgments}
J. Hu would like to thank Dr. Tingting Sun for the useful discussion on the calculation of $\Xi^-$ hypernuclei in relativistic mean field model. This work was supported in part by the National Natural Science Foundation of China (Grants No. 11375089, No. 11405090,  No. 11675083, and No. 11775119) and the Fundamental Research Funds for the Central Universities.

\end{document}